\documentclass[aps,pra,twocolumn,superscriptaddress,address,amsmath,showpacs]{revtex4}

\usepackage{graphicx}

\begin{document}


\title{Optimal thickness of rectangular superconducting microtraps 
for cold atomic gases }

\author{A. Markowsky}
\author{A. Zare}
\author{V. Graber}

\affiliation{Institut f\"ur Theoretische Physik and Center for 
Collective Quantum Phenomena, Universit\"at  T\"ubingen, Auf der Morgenstelle 14, D-72076 T\"ubingen, Germany}

\author{T. Dahm}

\affiliation{Institut f\"ur Theoretische Physik and Center for 
Collective Quantum Phenomena, Universit\"at  T\"ubingen, Auf der Morgenstelle 14, D-72076 T\"ubingen, Germany}

\affiliation{Universit\"at Bielefeld, Fakult\"at f\"ur Physik, 
Postfach 100131, D-33501 Bielefeld, Germany}

\date{\today}

\begin{abstract}
We study superconducting microtraps with rectangular shapes
for cold atomic gases. We present a general argument
why microtraps open, if brought close to the
surface of the superconductor. We show that for a given
width of the strips there exists an optimal thickness
under which the closest distance of the microtrap from the
superconductor can be achieved. The distance can be 
significantly improved, if the edge enhancement of the 
supercurrent near edges and corners is exploited.
We compare numerical calculations with results from 
conformal mapping and show that conformal mapping can
often give useful approximate results.
\end{abstract}


\maketitle

Microstructured magnetic traps are useful devices for the
storage and manipulation of ultracold atoms and
Bose-Einstein-Condensates \cite{FortaghRMP}.
In particular they allow studying the interaction of
ultracold atoms with solid-state surfaces. However, the
lifetime of the atomic clouds close to the surface
is limited by noise radiation from current fluctuations
in the conducting surfaces \cite{Jones,Rekdal}. Therefore, the use of
superconducting materials for such devices is being
investigated 
\cite{Nirrengarten,Mukai,Roux,CanoPRL,Dikovsky,Sokolovsky,CanoEPJD}.
The energy gap of a superconductor strongly suppresses 
noise in the relevant frequency range and thus allows
significantly longer lifetimes \cite{Skagerstam,Hohenester,Kasch,Fermani}.
Also, the use of superconductors promises the
construction of hybrid quantum systems and coupling
with superconducting electronic elements like Josephson Junctions,
SQUIDs, or microwave cavities as a possible route for
quantum computation \cite{Verdu,Folman}.

One drawback of superconducting microtraps is the strong redistribution
of current due to the Meissner effect.
Cano et al have shown that the Meissner effect in superconducting
microtraps lowers the trap depth and decreases the
trapping frequencies \cite{CanoPRL,CanoPRA} hampering the
close approach of the atomic cloud to the superconductor
surface. The closest distance reported so far in
a superconducting microtrap has been $20$~$\mu$m \cite{Kasch}.
In principle, trap depth and trapping frequencies can
always be increased with stronger currents and fields.
However, the maximum current is limited by the critical 
current density in a superconducting
material. As a possible way out of this difficulty it has
been suggested to utilize the vortex state of type-II
superconductors in which magnetic field partially penetrates
the superconductor and reduces the Meissner effect \cite{Emmert,Siercke}.
However, this reintroduces noise radiation coming
from the normal conducting cores of the vortices and
mitigates the use of the superconducting material \cite{Nogues}.

In this work, we study how the Meissner effect
lowers the magnetic trap depth. The screening current
due to the Meissner effect becomes particularly strong
near corners and edges of a superconducting structure
\cite{DS}. Here, we show that this edge 
enhancement of the screening current in a microtrap
formed by a microstrip of rectangular cross-section can be utilized
to bring a microtrap significantly closer to the
surface of the superconductor. This allows a
stronger atom-surface coupling without the need
to enter the vortex state. We perform calculations
of the current and field distributions in the vicinity
of the microstrip and show that there exists an
optimal strip thickness at which the closest distance
to the surface can be achieved.

In superconducting microtraps cold atoms with magnetic
moments are
caught in magnetic field minima. A magnetic field
minimum can typically be created by the inhomogeneous
field of a conducting wire and a homogeneous external
magnetic field perpendicular to the wire \cite{FortaghRMP}. 
In such a situation there will always be a magnetic field 
zero in the plane perpendicular to the wire.
There is a fundamental reason why magnetic microtraps
open in the vicinity of the surface of a superconductor,
which derives from Ampere's law and the Meissner effect:
the radial trapping frequency is proportional to the normal 
derivative of the tangential magnetic field component 
$\frac{\partial B_t}{\partial n}$ near the surface.
Due to Ampere's law we have 
$\frac{\partial B_t}{\partial n}= \frac{\partial B_n}{\partial t}$.
The normal component $B_n$ is continuous at a surface.
As $\vec{B}=0$ inside the superconductor, it follows
$\frac{\partial B_n}{\partial t}=0$ at the surface.
Therefore, the radial trapping frequency approaches zero
when a superconducting surface is approached and the trap
has to open. 

There are two ways out of this difficulty, which will be
studied in the following. First of all, the above argument
is true only on lengthscales larger than the
London penetration depth $\lambda$, because the magnetic
field inside the superconductor decreases continuously to
zero on that scale. Secondly, there can be high screening
current flowing in the superconductor, which may create
high local field gradients. This is particularly true
near corners and edges. In the following we will
study magnetic microtraps generated by a rectangular
superconducting (thin film) strip and investigate the behavior
near the corners.

\begin{figure}[t]
\includegraphics[width=0.87 \columnwidth]{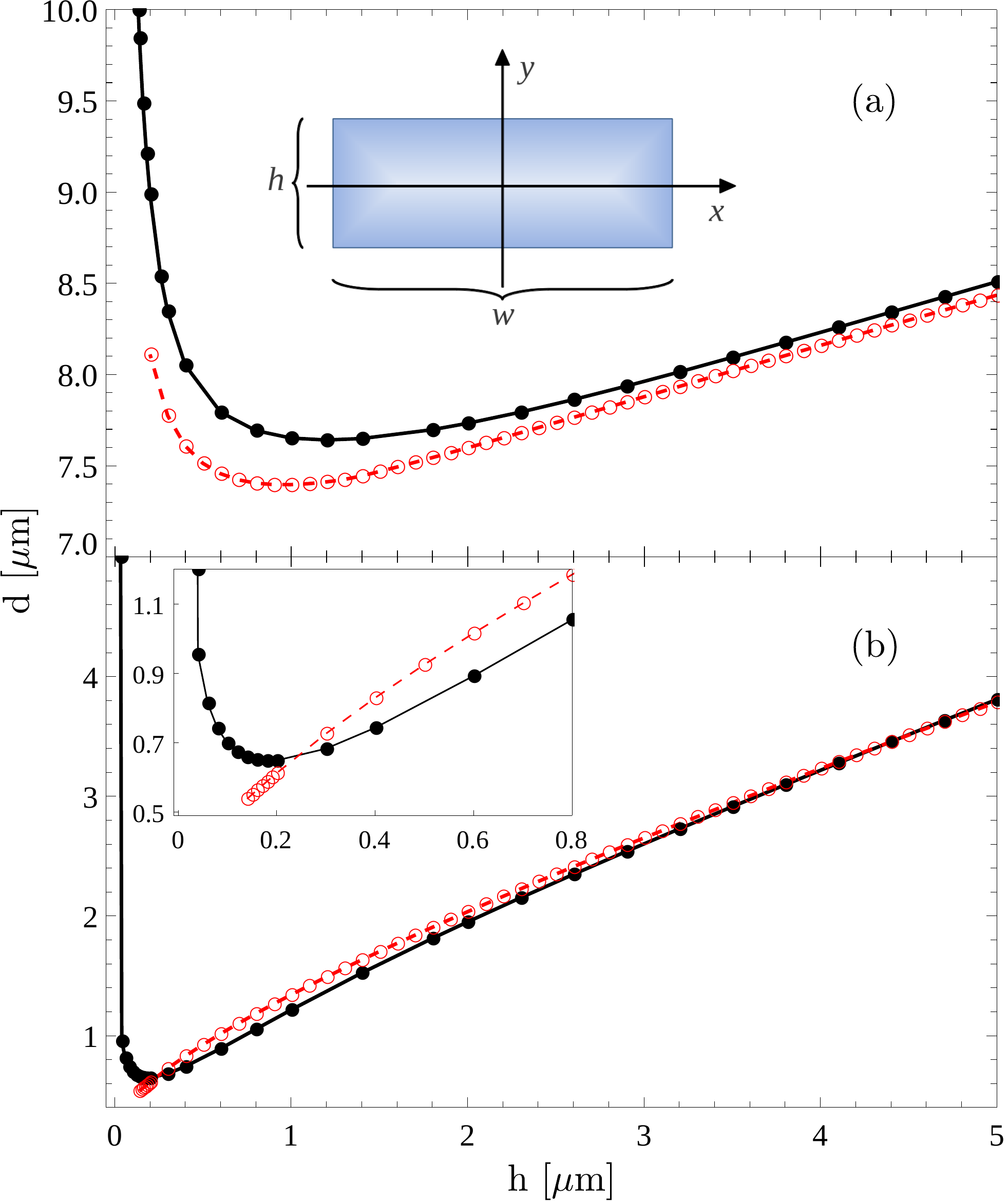}
\caption{\label{Fig01} (Color online)
Closest reachable distance between the trap and a strip 
of width $w=20$~$\mu$m as
a function of film thickness $h$ under the constaints
discussed in the main text. Black solid circles show 
the results of the numerical calculation, red open circles 
are the results obtained from conformal mapping.
a) The trap approaches the strip along the $y$-axis,
external magnetic field is applied along the $x$-axis.
Inset: Geometry of the strip. b) The trap approaches the 
strip along the $x$-axis,
external magnetic field is applied along the $y$-axis. 
The inset shows an enlarged region near the minimum.}
\end{figure}

The geometry is shown in the inset to Fig.~\ref{Fig01}~(a).
The cross-section of the strip lies in the $x$-$y$-plane
with a current $I$ flowing through in negative $z$-direction.
A constant external magnetic field $\vec{B}_{\rm ext}$ is
applied within the $x$-$y$-plane, such that a
magnetic field zero is created somewhere in the 
$x$-$y$-plane outside the strip. In addition, a small
offset field $\vec{B}_{\rm o}$ of 1~Gauss is applied in $z$-direction
in order to avoid Majorana spin-flip losses \cite{FortaghRMP}.
The offset field turns the magnetic field zero into a
magnetic field minimum which locally acts as
a harmonic oscillator potential for the atoms.
We calculate the current distribution in the
strip using the numerical method by Brandt and Mikitik \cite{EHB}.
In this method the electromagnetic kernel is discretized on
a grid for the strip cross-section and inverted in order
to obtain the current distribution inside the strip.
For the grid we choose an inhomogeneous mesh which is
more closely spaced near the corners of the strip \cite{EHB2,DS}.
After determination of the current distribution, the
magnetic field distribution is obtained from the Biot-Savart law.

For a given direction of $\vec{B}_{\rm ext}$ we vary
the total current $I$ and the magnitude $\left| \vec{B}_{\rm ext}
\right|$ in order to minimize the distance of the magnetic field 
minimum from the strip under the following two constraints:
1) the magnitude of the magnetic field at all points on the surface of the
superconductor is required to be larger than the trap minimum
by an amount of $10 k_B T / \mu_B$ in order to ensure that the
atoms cannot escape from the trap. In the calculations below
we choose $T=200$~nK, a typical value that can be
achieved in cold atom experiments. 2) The current density in the strip
is required to stay below the critical current density $j_c$
at all points. The critical current density generally depends
on the quality of the material. Here, we will take
$j_c=10^6$~A/cm$^2$ which is a typical value for
Nb thin films \cite{Huebener}.
Using this optimization procedure we finally determine the closest distance
of the trap minimum allowed under the two constraints.

In order to check our results and study the influence
of the penetration depth $\lambda$ we compare our 
numerical calculations with the method of conformal mapping \cite{Henrici}.
Conformal mapping allows calculation of the field distribution
and the surface current distribution semi-analytically for
polygonally shaped strips. This method is strictly
valid only for penetration depth $\lambda \rightarrow 0$.
However, it gives useful results as long as the trap
minimum is farther away from the superconductor than
the lengthscale $\lambda$, as can be seen from the
results below. We use the conformal map for a rectangular
strip that has been given by Chen et al \cite{Chen}.
Within the conformal mapping method the current density
diverges at the corners, which is a consequence
of the $\lambda \rightarrow 0$ limit. For our calculations
we need to regularize these divergencies. We do so
by requiring that the surface current density stays
below $\lambda j_c$ at a distance of order $\lambda$
from the corners as a replacement for the second 
constraint mentioned above.

In Fig.~\ref{Fig01} we show results for a $w=20$~$\mu$m wide
strip. The penetration depth was taken $\lambda=100$~nm,
a value appropriate for Nb thin films.
Fig.~\ref{Fig01}~(a) shows the closest reachable distance $d$ from
the strip as a function of film thickness $h$ for an external 
magnetic field applied in $x$-direction. In this case the
magnetic field minimum always lies on the negative
$y$-axis at the position $y_0$, i.e. $d=|y_0|-h/2$.
The black solid circles show the results of the numerical
calculation, while the red open circles are the results
obtained from conformal mapping.
Reducing the film thickness $h$ the reachable distance
initially decreases slightly and reaches a minimum
at a film thickness of $h=1.2$~$\mu$m with a distance of
7.64~$\mu$m from the strip. Further reduction of the
film thickness leads to a quick increase of the
closest distance. Both the numerical calculation and
the conformal mapping method yield similar results.
The conformal mapping method deviates from the
numerical calculation on the order of the
penetration depth and becomes invalid
for smaller strip dimensions, as expected. 

Qualitatively,
the appearance of an optimum film thickness can be understood
as follows: when the film thickness is large, the current
flows mainly at the surface of the strip within a layer of
thickness $\lambda$. For a fixed current density $j_c$ in
the strip the total current $I$ then varies with the
circumference of the strip.
Decreasing $h$ leads to a slight decrease of $I \propto 1 + h/w$.
If one wants to keep the trap minimum at the same position,
one has to decrease the magnitude of the external
magnetic field by the same factor $1 + h/w$. Then the
magnetic field at the surface of the strip will increase
because the external field has decreased while the field due
to the current roughly remained constant because of the
smaller thickness. In effect, the trap depth has increased
and this means that the trap minimum can be brought closer
to the strip. This explains the $\propto 1 + h/w$ 
behavior of the
minimum trap distance seen at larger values of $h$.
When $h$ becomes smaller, the current flow in the strip
becomes more homogeneous along the $y$-direction. Then
the total current decreases more quickly $I \propto h$.
In this case the trap depth decreases and the
trap minimum has to be brought farther away from
the strip in order to maintain the same trap depth.

\begin{figure}[t]
\includegraphics[width=0.87 \columnwidth]{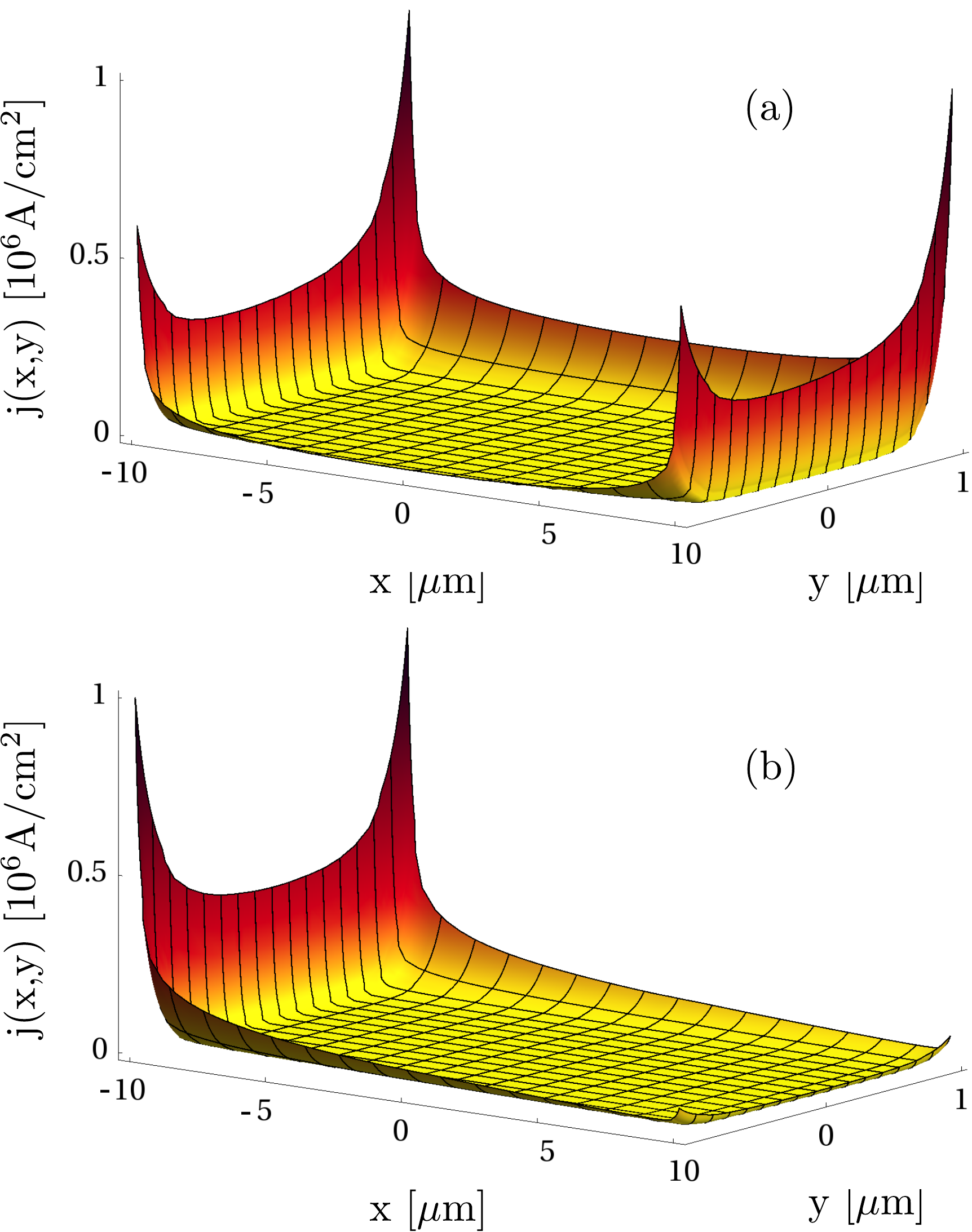}
\caption{\label{Fig02} (Color online)
Current distribution at optimized distance inside
a strip of width $w=20$~$\mu$m and thickness $h=2$~$\mu$m.
a) The external magnetic field is applied in
$x$-direction and the trap minimum lies on the
negative $y$-axis.
b) The external magnetic field is applied in
$y$-direction and the trap minimum lies on the
positive $x$-axis.}
\end{figure}

The situation changes completely, if the external
magnetic field is applied in $y$-direction and
the trap minimum approaches the strip from
its thin side along the $x$-axis, as shown in
Fig.~\ref{Fig01}~(b). In this situation
strong screening currents have to flow at the outer
edges of the strip in order to screen out the
external magnetic field from the inner part of the
strip. Correspondingly, there exist much stronger
field gradients in the vicinity of the trap minimum
and the reachable distance
initially decreases roughly proportional to $h$.
This behavior stops when a film thickness of $h \approx \lambda$
is reached, because then the magnetic field can
penetrate the strip and the screening is reduced.
As Fig.~\ref{Fig01}~(b) shows, the optimum trap
distance is reached at a film thickness of $h=0.19$~$\mu$m 
with a distance of 0.65~$\mu$m from the strip, 
about a factor of 12 times smaller than
in Fig.~\ref{Fig01}~(a). The conformal mapping
method does not show the increase of the trap
distance at low values of $h$. This is due to
the fact that in this range $h \sim \lambda$ the
conformal mapping method is not valid anymore.

\begin{figure}[t]
\includegraphics[width=0.87 \columnwidth]{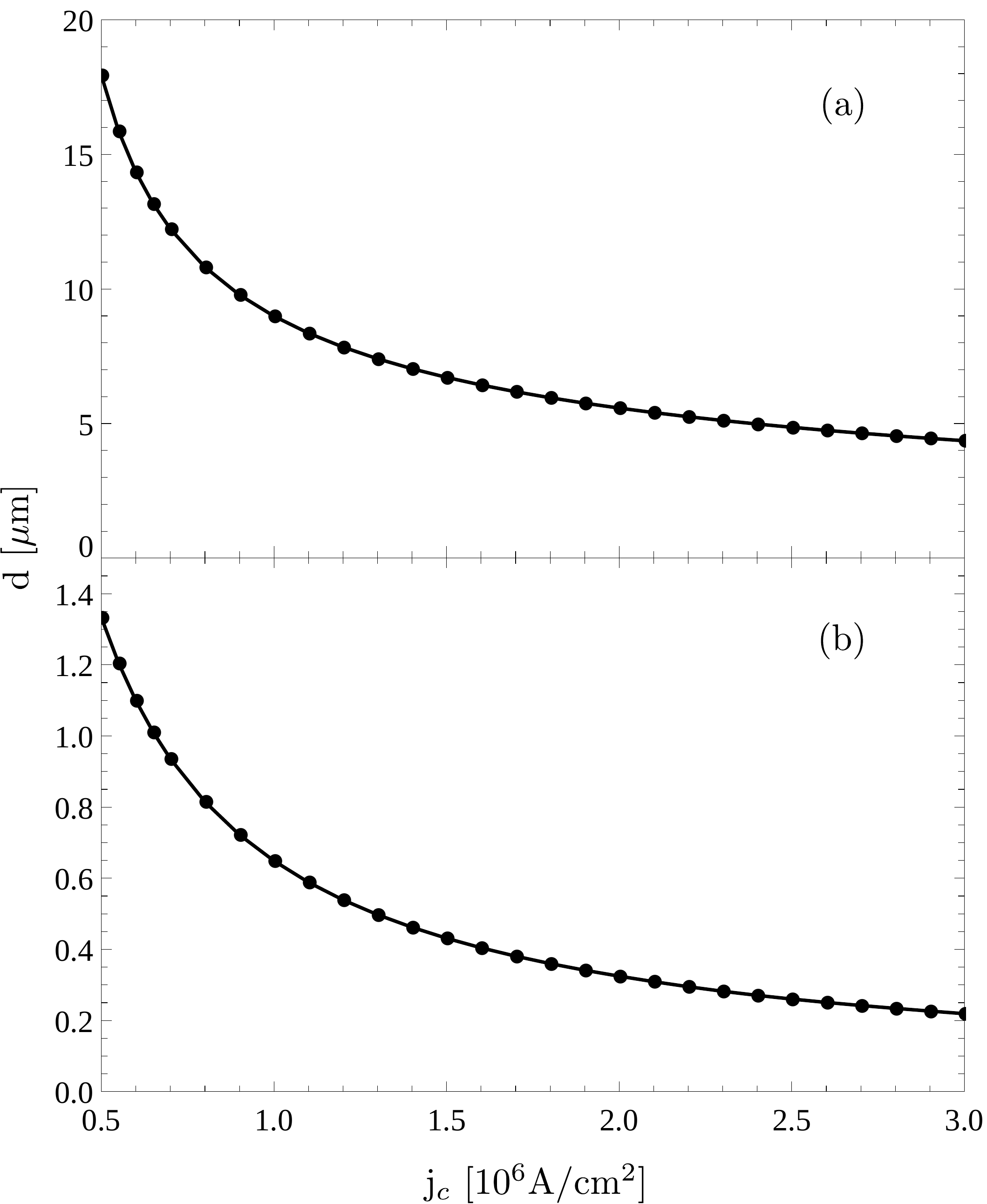}
\caption{\label{Fig03}
Optimum distance as a function of the
critical current density $j_c$ for a strip
with width $w=20$~$\mu$m and thickness $h=0.2$~$\mu$m. 
a) The external magnetic field is applied in
$x$-direction and the trap minimum lies on the
negative $y$-axis.
b) The external magnetic field is applied in
$y$-direction and the trap minimum lies on the
positive $x$-axis.}
\end{figure}

In Fig.~\ref{Fig02} we show the current distribution
inside a strip of width $w=20$~$\mu$m and thickness 
$h=2$~$\mu$m under optimized conditions.
Fig.~\ref{Fig02}~(a) shows the current distribution
when the trap minimum approaches the strip along
the negative $y$-axis. In this case, the current flows
mainly along the three edges of the strip away from
the trap minimum. There is only small current flowing
at the edge closest to the trap minimum. This is clear,
because under optimized conditions the magnetic field
becomes very small close to this edge and only small
screening current has to flow here. The highest
current density ($j_c$) is found at the two corners opposite
to the trap.
Fig.~\ref{Fig02}~(b) shows the current distribution
when the trap minimum approaches the strip along
the $x$-axis. In this situation, the current flows
mainly along the short edge of the strip opposite to
the trap minimum. The current along the long edges of
the strip can almost be neglected. This is due to the
fact that the external magnetic field is applied
along $y$-direction, which creates a strong screening current
at the short outer edges of the strip \cite{EHB}. 
The highest current density is 
again found at the two corners opposite to the trap.
The magnetic field gradient in the trap center is
$4.5$~T/m for the case in Fig.~\ref{Fig02}~(a) and
$9.9$~T/m for Fig.~\ref{Fig02}~(b).

The optimum distance that can be achieved under
the two constraints above obviously depends
on the quality of the material via the
critical current density $j_c$. In Fig.~\ref{Fig03}
we demonstrate how the optimum distance
varies with $j_c$ for a strip with width 
$w=20$~$\mu$m and thickness $h=0.2$~$\mu$m
along the two coordinate axes.
For $j_c=3 \cdot 10^6$~A/cm$^2$, for example, 
one may approach
the strip $4.5$~$\mu$m from the $y$-axis and
$0.22$~$\mu$m from the $x$-axis. 

\begin{figure}[t]
\includegraphics[width=0.87 \columnwidth]{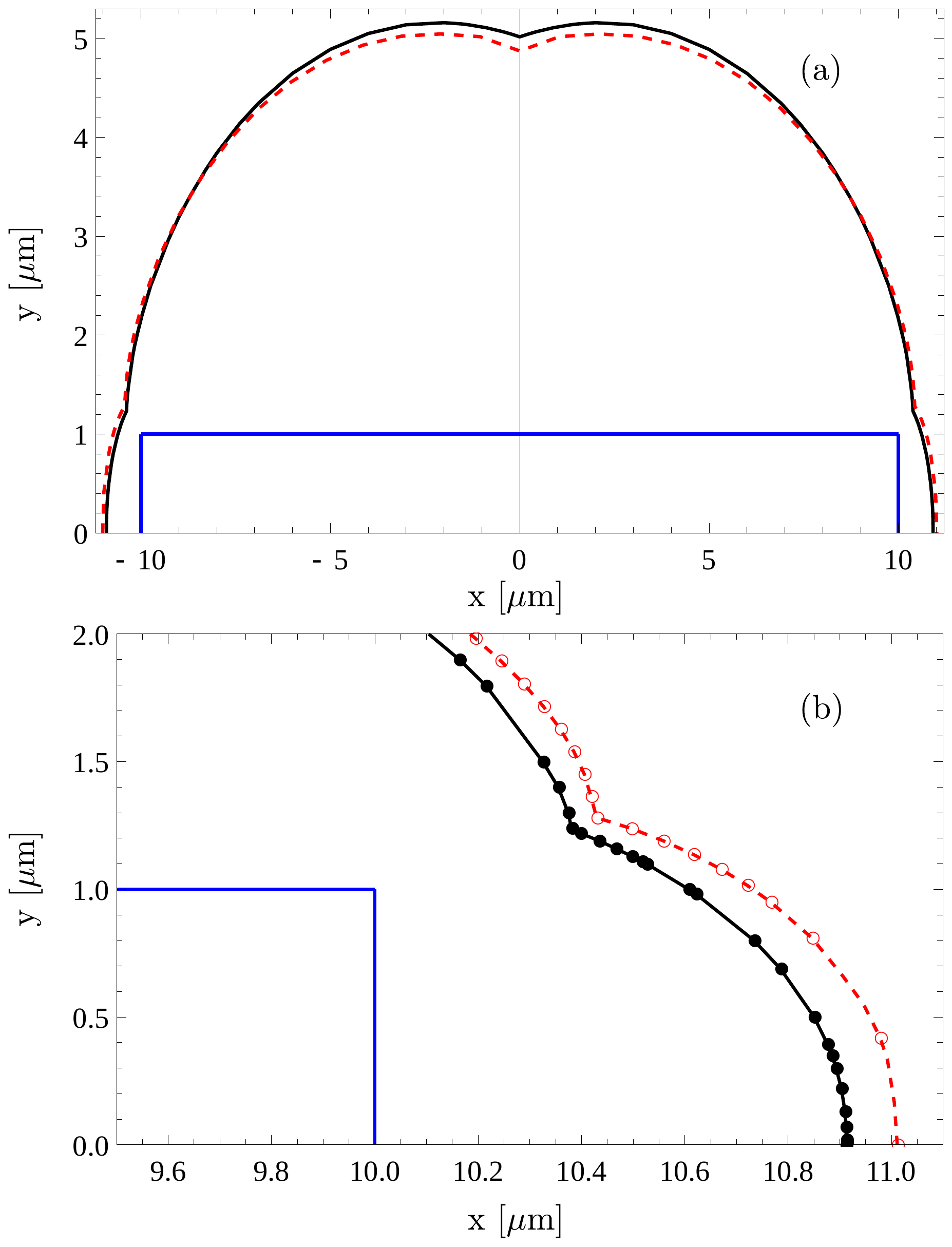}
\caption{\label{Fig04} (Color online)
a) Locus of points
of optimum distance for different directions
of the external magnetic field. Here, a strip
with width $w=20$~$\mu$m, thickness $h=2$~$\mu$m,
and $j_c=3 \cdot 10^6$~A/cm$^2$
was taken. Black solid lines (solid circles) show 
the results of the numerical calculation, red dashed lines 
(open circles)
are the results obtained from conformal mapping.
b) shows an enlarged region near the corner of the strip.}
\end{figure}

In practice, an approach from the thin side of the
strip may not always be feasible. Nevertheless,
the edge enhancement of the screening current
can still be exploited, if the trap is brought
close to the corner of the strip.
In Fig.~\ref{Fig04} we show the locus of points
of optimum distance for different directions
of the external magnetic field for a strip
with width $w=20$~$\mu$m and thickness $h=2$~$\mu$m.
Both the results from the numerical calculation
and conformal mapping are shown. The results
differ only by the order of the penetration length,
as expected.
As one can see, the optimum distance lies even
closer to the strip in the vicinity of the corner.
This is related to the stronger $l^{-1/3}$ divergence
of the surface current in the corners in the
limit $\lambda \rightarrow 0$, where
$l$ is the distance from the corner \cite{EHB}.

To conclude, we have shown that for a 
rectangular superconducting strip of width $w$
there exists an optimal film thickness for
which a microtrap can be brought closest to
the surface of the superconductor. The closest
approach can be achieved near the corners or
short edges of the strip. This is due to the
edge enhancement of the screening current.
Near the corners distances in the sub-micrometer
range can be realistically achieved. Such distances
are essential for good control coupling in
quantum computation \cite{Folman}, for coupling 
of atoms with a superconducting qubit via the near-field 
of a microwave cavity mode \cite{Verdu}, or
for studying the Casimir-Polder force.

We acknowledge useful discussions with D.~Cano and J.~Fortagh.
This work was supported by the Deutsche Forschungsgemeinschaft
via SFB/TR21.


\begin{thebibliography}{99}

\bibitem{FortaghRMP} J.~Fortagh and C.~Zimmermann, {Rev. Mod. Phys.}
{\bf 79}, 235 (2007).

\bibitem{Jones} M.P.A.~Jones, C.J.~Vale, D.~Sahagun, B.V.~Hall, and 
E.A.~Hinds, Phys. Rev. Lett. {\bf 91}, 080401 (2003).

\bibitem{Rekdal} P.K.~Rekdal, S.~Scheel, P.L.~Knight, and E.A.~Hinds , 
{Phys.~Rev.~A} {\bf 70}, 013811 (2004).

\bibitem{Nirrengarten} T.~Nirrengarten et al, {Phys. Rev. Lett.} {\bf 97}, 
200405 (2006).

\bibitem{Mukai} T.~Mukai et al,  {Phys. Rev. Lett.} {\bf 98}, 
260407 (2007).

\bibitem{Roux} C.~Roux et al, {EPL} {\bf 81}, 56004 (2008).

\bibitem{CanoPRL} D.~Cano et al, {Phys. Rev. Lett.} {\bf 101}, 
183006 (2008).

\bibitem{Dikovsky} V.~Dikovsky et al, V.~Sokolovsky, B.~Zhang, C.~Henkel, 
and R. Folman, {Eur. Phys. J. D} {\bf 51}, 247 (2009).

\bibitem{Sokolovsky} V.~Sokolovsky, L.~Prigozhin,
and V.~Dikovsky, {Supercond. Sci. Techn.} {\bf 23}, 065003 (2010).

\bibitem{CanoEPJD} D.~Cano et al, {Eur.~Phys.~J.~D} {\bf 63}, 
17 (2011).

\bibitem{Skagerstam} B.-S.K.~Skagerstam, U.~Hohenester, A.~Eiguren,
and P.K.~Rekdal, {Phys. Rev. Lett.} {\bf 97}, 070401 (2006).

\bibitem{Hohenester} U.~Hohenester, A.~Eiguren, S.~Scheel, E.A.~Hinds, 
{Phys.~Rev.~A} {\bf 76}, 033618 (2007).

\bibitem{Kasch} B.~Kasch et al, New~J.~Phys. {\bf 12}, 065024 (2010).

\bibitem{Fermani} R.~Fermani et al, J.~Phys.~B {\bf 43}, 095002 (2010).

\bibitem{Verdu} J.~Verdu et al, {Phys. Rev. Lett.} {\bf 103}, 
043603 (2009); G.~Bensky et al, Quantum Inf. Process. {\bf 10}, 1037 (2011).

\bibitem{Folman} R.~Folman, Quantum Inf. Process. {\bf 10}, 995 (2011);
R.~Salem et al, New~J.~Phys. {\bf 12}, 023039 (2010).

\bibitem{CanoPRA} D.~Cano et al, {Phys.~Rev.~A} {\bf 77},
063408 (2008).

\bibitem{Emmert} A.~Emmert et al, {Phys.~Rev.~A} {\bf 80},
061604 (2009).

\bibitem{Siercke} M.~Siercke et al, {Phys.~Rev.~A} {\bf 85},
041403(R) (2012); B.~Zhang et al, {Phys.~Rev.~A} {\bf 85},
013404 (2012).

\bibitem{Nogues} G.~Nogues et al, EPL {\bf 87}, 13002 (2009).

\bibitem{DS} T.~Dahm and D.~J.~Scalapino, {J.~Appl.~Phys.} {\bf 81},
2002 (1997).

\bibitem{EHB} E.~H.~Brandt and G.~P.~Mikitik,  {Phys. Rev. Lett.} {\bf 85}, 
4164 (2000).

\bibitem{EHB2} E.~H.~Brandt, {Phys. Rev. B} {\bf 58}, 
6506 (1998).

\bibitem{Huebener} R.~P.~Huebener et al, {J.~Low~Temp.~Phys.} {\bf 19},
247 (1975).

\bibitem{Henrici} P.~Henrici, {\it Applied and Computational Complex
  Analysis}, Vol.~I, Wiley (1988).

\bibitem{Chen} D.-X.~Chen, C.~Prados, E.~Pardo, A.~Sanchez,
and A.~Hernando, {J.~Appl.~Phys.} {\bf 91},
5254 (2002).



\end{thebibliography}
\end{document}